# Analytical Correlation in the H₂ Molecule from the Independent Atom Ansatz

Alanna "Lanie" Leung, Alexander V. Mironenko*

*Department of Chemical and Biomolecular Engineering,*
*University of Illinois Urbana-Champaign, Urbana, Illinois 61820*

*Email: alexmir@illinois.edu

**Abstract:** The independent atom ansatz of density functional theory yields an accurate analytical expression for dynamic correlation energy in the H₂ molecule: $E_c = 0.5(1 - \sqrt{2})(ab|ba)$ for the atom-additive self-consistent density $\rho = |a|^2 + |b|^2$. Combined with exact atomic self-exchange, it recovers more than 99.5% of nearly exact SCAN exchange-correlation energy at $R > 0.5$ Å, differing by less than 0.12 eV. The total energy functional correctly dissociates the H-H bond and yields absolute errors of 0.002 Å, 0.19 eV, and 13 cm$^{-1}$ relative to experiment at the tight binding computational cost. The chemical bond formation is attributed to the asymptotic Heitler-London resonance of quasi-orthogonal atomic states ($-(ab|ba)$) with no contributions from kinetic energy or charge accumulation in the bond.

Accurate electronic correlation energy calculations are essential for predicting thermochemistry and kinetics across chemistry, catalysis, and materials science. As the many-body electronic structure problem is not analytically solvable, a variety of sophisticated approximations have been devised, such as finite-basis full configuration interaction (FCI) [1], coupled cluster methods [2-13], Møller-Plesset perturbation theory [14-19], correlation functionals of density functional theory (DFT) [20-41], and others. It is generally assumed [42] that more accurate correlation energies require more complicated mathematical objects that are more computationally costly to evaluate.

In this Letter, we demonstrate that for a prototypical H₂ molecule, it is possible to obtain a surprisingly simple and accurate analytical form of the dynamic correlation energy, using a redefined DFT reference state. A procedure is reported leading to the expression $E_c = 0.5(1 - \sqrt{2})(\varphi_1\varphi_2|\varphi_2\varphi_1)$, which holds for the atom-additive electron density $\rho = |\varphi_1|^2 + |\varphi_2|^2$. Here, $\varphi_a$ with $a$ = 1 or 2 are 1s orbitals with self-consistent exponents. The mathematical form of $E_c$ is obtained and assessed using the independent atom ansatz of density functional theory (DFT), introduced by one of us, which forms the basis for the nonempirical tight binding theory (NTB) [43]. Next, we discuss numerical results followed by the theory and its implications on the nature of the covalent H-H bond.

Figure 1 shows $\Delta E(R)$ binding energy curves for H₂, computed using NTB with analytical correlation, FCI, and DFT at PBE [32] and SCAN [38] levels of theory. NTB correctly predicts the binding energy curve shape throughout the tested range of internuclear distances and exhibits the correct bond dissociation behavior, in contrast to the restricted Kohn-Sham DFT. The correct dissociation is attributed to the absence of the spurious interaction of electrons with fractional spins on atoms [44,45]. The equilibrium bond lengths ($R_0$), bond energies ($\Delta E_0$), and wavenumbers ($\nu$) are accurately predicted with NTB, with absolute errors of 0.002 Å, 0.189 eV, and 13.3 cm$^{-1}$, respectively, relative to experiment (Table 1). Notably, NTB with the minimal basis and analytical correlation outperforms large-basis DFT/PBE in all three metrics for H$_2$. The remaining error in $\Delta E_0$ is attributed to the neglect of polarization (0.08 eV) and to the simplicity of the dynamic correlation expression, which omits interactions of 1s with $nlm$ states for $n > 1$ (0.11 eV, Section S9).

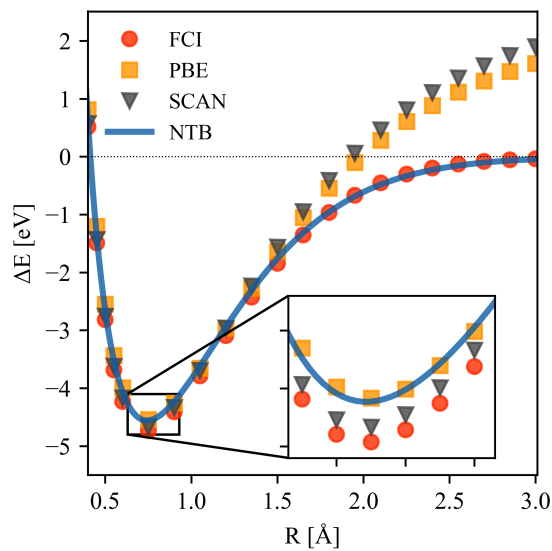

FIG. 1. H₂ binding energy curves of NTB with analytical correlation (blue solid line), restricted DFT/PBE (yellow squares) and DFT/SCAN (grey triangles), and FCI (red circles). DFT and FCI employ the cc-pVQZ basis set. NTB employs the self-consistent $\zeta$-STO basis set.

**Table 1.** H$_2$ equilibrium geometries ($R_0$), energies ($\Delta E_0$), and vibrational wavenumbers ($\nu$). Experimental data are taken from Ref. [46,47].

| Method | $R_0$ (Å) | $\Delta E_0$ (eV) | $\nu$ (cm$^{-1}$) |
|---|---|---|---|
| NTB | 0.743 | -4.558 | 4388 |
| PBE | 0.750 | -4.541 | 4312 |
| SCAN | 0.741 | -4.669 | 4425 |
| FCI | 0.742 | -4.732 | 4403 |
| Exp. | 0.741 | -4.747 | 4401 |

Figure 2 presents a comparison between the analytical $E_{xc}^{NTB}$ and the Kohn-Sham DFT exchange-correlation (XC) energies $E_{xc}^{KS}$, computed at exchange-only local density approximation (XLDA), generalized gradient approximation (GGA; PBE), and meta-GGA (SCAN) levels of theory. $E_{xc}^{NTB}$ is computed by combining analytical $E_c$ with the exact atomic self-exchange ($2 \times (-5\zeta/16)$); $E_{xc}^{KS}$ values are corrected for the static correlation error by adding $2 \times (E_{xc}^{KS}[1,0] - E_{xc}^{KS}[0.5,0.5])$, where the bracketed values correspond to atomic spin occupancies $[\alpha, \beta]$. Finally, the XLDA curve is shifted to match the exact exchange of free atoms. The difference between the analytical and nearly exact SCAN XC energy is less than 0.12 eV at $R > 0.5$ Å. XLDA, GGA, and meta-GGA values progressively approach the analytical line, suggesting that the increasingly complex XC functionals are capturing an intrinsically simple analytical form. It is notable that the local exchange in these methods accounts for the majority of interactions classified as dynamic correlation in NTB (*vide infra*).

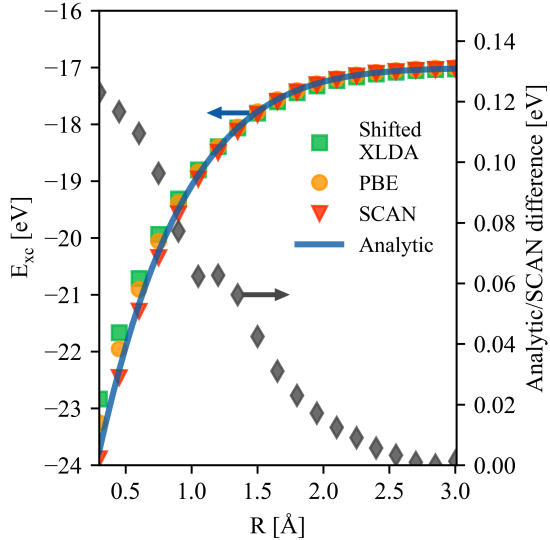

FIG. 2. Exchange-correlation energy of the H$_2$ molecule for the self-consistent density $\rho = |\varphi_1|^2 + |\varphi_2|^2$, computed analytically and using static-correlation-corrected (shifted) XLDA, PBE, and SCAN functionals.

NTB electron density matches closely FCI or Kohn-Sham DFT density in the vicinity of nuclei, exhibiting slightly positive deviations along the bond axis at $x < x_1$ and $x > x_2$, where $x_1$ and $x_2$ are nuclear coordinates, and very negative deviations at $x_1 < x < x_2$ (Fig. 3). We attribute the deviations to the absence of polarization functions in the $\zeta$-STO basis used for NTB, rendering atomic density contributions too isotropic (Fig. 3b). Significant density differences in the crucial inter-atomic region have a surprisingly minor effect on the energy accuracy. This observation echoes the general tendency among DFT functionals [48] and can be corroborated in the context of the first Theophilou theorem [49], stating that spherical densities around each nucleus (such as rescaled 1s orbitals in the present case) are sufficient to uniquely determine the ground state of the system. The effect of orbital rescaling amounts to bringing the cusp density closer to the exact value (Fig. 3c) to approximately satisfy the Kato cusp theorem [50] and thus completely specify the Hamiltonian.

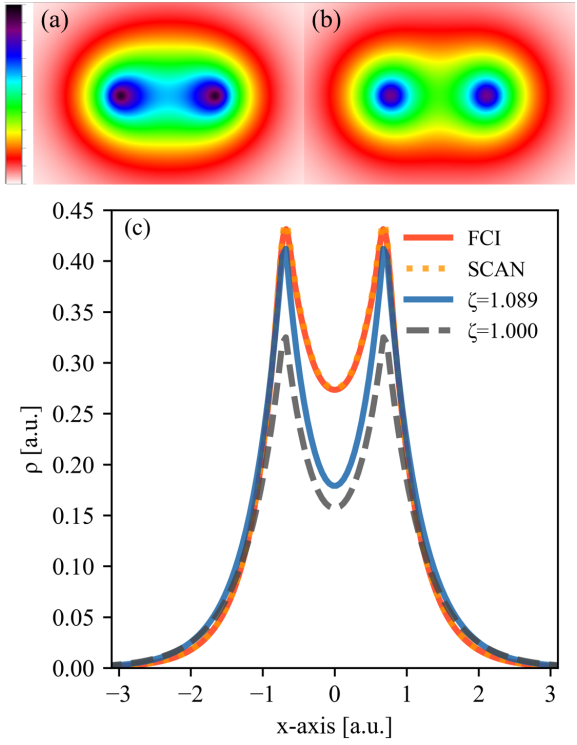

FIG. 3. Densities of $H_2$ with $R = 0.741$ Å obtained with FCI (a) and NTB (b). (c) compares FCI (red solid line), SCAN (yellow dotted line), and NTB density profiles along the bond axis. NTB profiles are reported for $\zeta$ optimized to 1.089 (blue solid line) and fixed to its free atom value of 1 (grey dashed line).

The numerical results for the $H_2$ molecule are obtained using the NTB total energy functional that implements the independent atom (IA) ansatz of DFT [43]:

$$E[\rho_1, \rho_2] = \sum_{a=1}^{2} E'_a + E_{es}^{\circ\circ} + E_{xc}^{\circ\circ}, \quad (1)$$

where $E'_a$ is the energy of a quasi-perturbed H atom, described below; $E_{es}^{\circ\circ}$ is the inter-atomic electrostatic energy for $\rho = \rho_1 + \rho_2$; and $E_{xc}^{\circ\circ}$ is the inter-atomic XC energy in the IA ansatz. Atomic density is defined as $\rho_a = |\varphi_a|^2$, where the atom-localized states $|\varphi_a\rangle$ are referred to as atomions [43]. The IA-XC functional $E_{xc}^{\circ\circ}$ is decomposed as

$$E_{xc}^{\circ\circ} = -2S\beta + 2\beta + E_c^{\leftrightarrow}, \quad (2)$$

where the terms correspond to orthogonalization (steric repulsion), hybridization (resonance), and dynamic correlation, respectively. $E_c^{\leftrightarrow}$ along with $\sum_a E_{xc}[\rho_a]$ make up the XC energy in Kohn-Sham DFT [43]. $S$ is the overlap integral, and $\beta \leq 0$ is the off-diagonal element (resonance integral) of the 2x2 Hückel matrix $\mathbf{D}$. According to the exchange-static correlation interpretation of the hybridization energy [43], $\beta$ is defined as

$$\beta = D_{ab} = D_{ba} = -\frac{1}{2}(\varphi_a\varphi_b|\varphi_a\varphi_b). \quad (3)$$

The non-local $\beta$ in eq. (3) exceeds in accuracy the local $\beta$ introduced in Ref. [51] (Section S3). Reconciliation of non-local and local NTB approaches is discussed in Section S8.

The $|\varphi_a\rangle$ states are the solutions to the *atomion equation*:

$$\varepsilon_a|\varphi_a\rangle = \left[-\frac{1}{2}\nabla^2 + v_a + v_b + j_b \right. \\ \left. +\mu_{xc}^{\leftrightarrow,a} - P_b(\mu_{x,ba}^{R\gg 0} + \eta_{sc,ba}^{R\gg 0})\right]|\varphi_a\rangle, \quad (4)$$

where $P_b = |\varphi_b\rangle\langle\varphi_b|$, $v_b$ and $j_b$ are the $b$-nucleus and $\rho_b$ potentials, $\mu_{x,ba}^{R\gg 0} = \eta_{sc,ba}^{R\gg 0} = -0.25\varphi_a(\mathbf{r})\varphi_b(\mathbf{r}')/|\mathbf{r}-\mathbf{r}'|$ are the asymptotic inter-atomic exchange and static correlation operators, and $\mu_{xc}^{\leftrightarrow,a}$ is the inter-atomic XC potential. It is defined as [43]

$$\mu_{xc}^{\leftrightarrow,a} = \frac{\delta(E_c^{\leftrightarrow} + E_{hyb})}{\delta\rho_a}, \quad (5)$$

where $E_{hyb} = 2\beta = D_{ab} + D_{ba}$. In this study, eq. (4) is solved in the integral form through application of the virial theorem $\varepsilon_a = -\zeta^2/2$. Virial and variational approaches are compared in terms of performance in Section S4.

The NTB theory [43] is based on six principles (Section S2), which are employed herein to obtain analytical forms of $E_c^{\leftrightarrow}$ and $\mu_{xc}^{\leftrightarrow,a}$. To get $E_c^{\leftrightarrow}$ in $H_2$, we first recognize the asymptotic correspondence between NTB and FCI in the $R \gg 0$ limit, valid to $O(S^0)$ [43]:

$$E_{xc}^{\infty} = E_{hyb} + E_c^{\leftrightarrow} = E_x^{\infty}|_{HF} + E_c^{\infty}|_{CI}, \quad (6)$$

where $E_x^{\infty}|_{HF} = E_c^{\infty}|_{CI} = -0.5(\varphi_1\varphi_2|\varphi_2\varphi_1)$ are the asymptotic inter-atomic exchange and correlation energies [43]. At very large $R$, electrons must become localized on atoms and distinguishable, and thus the inter-atomic exchange associated with indistinguishability (one-half of $D_{ab}$) must disappear (Principle 2, Section S2) [43]. To satisfy the asymptotic correspondence with valence bond (VB) theory yielding $E_c^{\infty}|_{VB} = -(\varphi_1\varphi_2|\varphi_2\varphi_1)$, $E_{xc}^{\infty}$ in eq. (6) must then contain an extra term $K_{12}^{\leftrightarrow} = -0.5(\varphi_1\varphi_2|\varphi_2\varphi_1)$, which, we argue, arises from $E_c^{\leftrightarrow}$. Using the asymptotic correspondence with FCI and the formal completeness of the atomion basis, we arrive at $E_c^{\leftrightarrow}$ that resembles minimal-basis FCI correlation [52]:

$$E_c^{\leftrightarrow} = \Delta_{NTB} - (\Delta_{NTB}^2 + (K_{12}^{\leftrightarrow})^2)^{1/2}. \quad (7)$$

To motivate an expression for $\Delta_{NTB}$, we first note that its FCI equivalent is $(h_{22} - h_{11}) + (J_{22} - J_{11}) - 0.5(K_{22} - K_{11})$ [52], where the indices correspond to bonding and antibonding MOs, and $h, J$, and $K$ are the standard one-electron, Hartree, and exchange integrals, respectively. In the $R \gg 0$ limit, both MOs have $\pm 1/\sqrt{2}$ coefficients and identical densities, and thus all intra-atomic terms and $J$ integrals cancel out, while inter-atomic contributions to $h_{ii}$ are 0 to $O(S^0)$. The inter-atomic parts of $K_{ii}$ ($-0.5(\varphi_1\varphi_2|\varphi_2\varphi_1)$) are asymptotically local (Principle 2) and at $R \gg 0$ are equivalent to $(-1)^{i-1}\langle\varphi_1|\mu_{x,i}(\mathbf{r})|\varphi_2\rangle$ for $i = 1, 2$. As the bonding and antibonding densities are identical in the same limit, so are $\mu_{x,i}(\mathbf{r})$'s. After returning the system from asymptotic to chemically bonding $R$ and replacing local integrals by their non-local counterparts ($\pm 0.5(\varphi_1\varphi_2|\varphi_2\varphi_1)$; Principle 3), we obtain

$$\Delta_{NTB} = \frac{1}{2}(\varphi_1\varphi_2|\varphi_2\varphi_1)$$
$$= -(D_{12}^x + D_{21}^x) = \frac{\varepsilon_2^{mo} - \varepsilon_1^{mo}}{2}, \quad (8)$$

where $\varepsilon_1^{mo}$ and $\varepsilon_2^{mo}$ are eigenvalues (molecular orbital energies) of the $\mathbf{D}$ Hückel matrix, and $D_{12}^x = D_{21}^x$ are the exchange fractions of $D_{12} = D_{21}$. After substitution of $\Delta_{NTB}$ from eq. (8) and $K_{12}^{\leftrightarrow} = -0.5(\varphi_1\varphi_2|\varphi_2\varphi_1)$ into eq. (7), a very simple analytical form of the dynamic correlation energy is obtained:

$$E_c^{\leftrightarrow} = \frac{1-\sqrt{2}}{2}(\varphi_1\varphi_2|\varphi_2\varphi_1). \quad (9)$$

It is notable that $E_c^{\leftrightarrow}$ formally corresponds to an unconventional *one-particle* CI theory, where $\varepsilon_1^{mo}$ and $\varepsilon_2^{mo}$ take place of ground and excited state energies, and both electrons occupying the bonding orbital act as a single effective particle with energy $\varepsilon_1^{mo}$, akin to the Cooper pair. According to this interpretation, $E_c^{\leftrightarrow} = 0$ in the triplet H$_2$, which is confirmed semi-quantitatively in Section S7. Eq. (9) is reconciled with the asymptotic VB theory in Section S8 through the analysis of quantum effects.

To obtain an analytical form of $\mu_{xc}^{\leftrightarrow,a}$ using eq. (5), we first note that the derivative of the static correlation part of $D_{ab}$ must be taken as zero [43]. The derivatives of its exchange fraction and of $E_c^{\leftrightarrow}$ are taken at $R = R_\infty$ ($R_\infty$ is a very large finite number) after $\mu_{x,ab}^{R\gg0} \to \mu_{x,ab}^{R\gg0} + \delta$ substitution in $D_{ab}^x = \langle\varphi_a|\mu_{x,ab}^{R\gg0}|\varphi_b\rangle$, where $\delta$ accounts for quantum effects that become significant at $R \gg 0$ [43]. In that limit, the resonance integrals are recognized as asymptotically local (Principle 2): $D_{ab}^x \to -\sqrt{\pi}\langle\varphi_a|\varphi_a|\varphi_b\rangle$ and $D_{ba}^x \to -\sqrt{\pi}\langle\varphi_b|\varphi_b|\varphi_a\rangle$. From the orbital parity rule (Principle 4) it follows that $D_{ab}^x$ is a functional of $\rho_a = |\varphi_a|^2$ and thus enters $\mu_{xc}^{\leftrightarrow,a}$, whereas $D_{ba}^x$ is not, since it contains an odd number of $\varphi_a$ orbitals. After taking the derivative, returning from the asymptotic limit to chemically bonding $R$ values, and adopting non-local resonance integral forms, a simple expression for the $\langle\varphi_a|\mu_{xc}^{\leftrightarrow,a}|\varphi_a\rangle$ integral is obtained:

$$\langle\varphi_a|\mu_{xc}^{\leftrightarrow,a}|\varphi_a\rangle = -\frac{\sqrt{2}}{4}(\varphi_a\varphi_b|\varphi_b\varphi_a). \quad (10)$$

It is notable that $E_c^{\leftrightarrow}$ and $\mu_{xc}^{\leftrightarrow,a}$ prefactors in eq. (9) and (10) are not linked by the $\mu_{xc} = \delta E_{xc}/\delta\rho$ relationship at $R < R_\infty$, and $\mu_{xc}^{\leftrightarrow,a}$ is a functional derivative only asymptotically at $R = R_\infty$ (Principle 6). Eq. (10) is further validated through the numerical performance comparison of various $\mu_{xc}^{\leftrightarrow,a}$ prefactors in Section S6.

The last remaining element of the NTB theory is the energy $E'_a$ of a quasi-perturbed atom $a$, which formally corresponds to $\rho_a = |\varphi_a|^2$. To obtain its form, we first introduce the promotion energy through $E'_a = E[\rho^0_a] + E_{prom}$, where $\rho^0_a = |\varphi^0_a|^2$ is density of a free H atom. Then, we note that in $\varphi_a = \varphi^0_a + \delta\varphi_a$, $\delta\varphi_a$ has the same degree of smallness as overlap $S$. Since NTB equations were derived by retaining terms up to $O(S)$ [43], $E_{prom}$ can only contain up to linear terms in $\delta\varphi_a$. It is easy to show that $E_{prom} = 2\varepsilon^0_a \langle \delta\varphi_a | \varphi^0_a \rangle$, where $\varepsilon^0_a = -0.5$. Since $\langle \varphi_a | \varphi_a \rangle = \langle \varphi^0_a | \varphi^0_a \rangle = 1$, we conclude that

$$E_{prom} = 0. \quad (11)$$

$E_{prom} = 0$ and thus $E'_a = E[\rho^0_a]$ are equivalent to replacing $\varepsilon_a \to \varepsilon^0_a$ while retaining the perturbed $\varphi_a$ in evaluations of all inter-atomic terms in eq. (2). This remarkable effect is referred to as the *atomion renormalization*.

The H$_2$ NTB binding energy expression with analytical dynamic correlation takes a very simple form:

$$\Delta E(R) = E^{\infty}_{es}$$
$$+ \left( -\frac{1+\sqrt{2}}{2} + \langle \varphi_1 | \varphi_2 \rangle \right) (\varphi_1 \varphi_2 | \varphi_2 \varphi_1), \quad (12)$$

The $-0.5(1+\sqrt{2})$ prefactor can be split into exchange $(-0.5)$, static correlation $(-0.5)$, and dynamic correlation $(0.5(1-\sqrt{2}))$ contributions. For quantitative accuracy, it is crucial to optimize (contract) $\varphi_1$ and $\varphi_2$ as a function of $R$ (Section S5) according to eq. (S3). Further implementation details along with the source code example are provided in Sections S1 and S10.

Eq. (12) is uniquely suited to obtain insights into the nature of chemical bonding. Historically, the chemical bond was explained as being mainly electrostatic in nature [53-59] and associated with the interatomic electron density buildup. Later, it was shown that kinetic energy lowering plays a central role in bonding [60-67]. However, quantitative analysis by Levine and Head-Gordon [68] demonstrated that the stabilizing kinetic energy effect is only restricted to hydrogen-containing bonds. In contrast to these earlier findings, eq. (12) indicates that within the IA ansatz of DFT, the kinetic energy makes no contribution to bonding in H$_2$! This observation seemingly contradicts the virial theorem, and in Section S11 we discuss a possible mechanism for this effect. Fig. 3 also shows that the electron density buildup in the interatomic region is not essential for capturing chemical bonding.

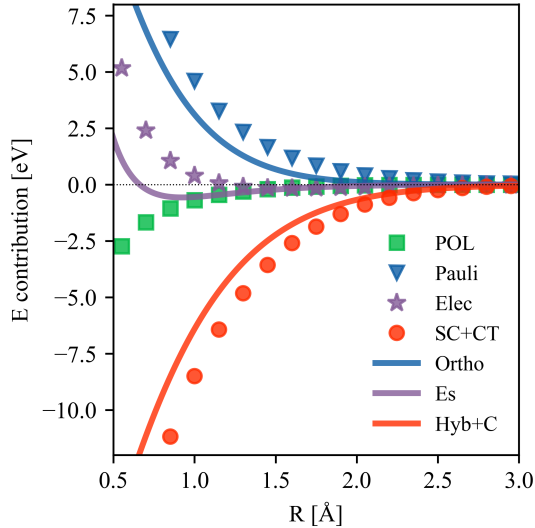

FIG. 4. Decomposition of the H$_2$ binding energy within NTB (lines) and using the ALMO-EDA method for DFT (points).

In Figure 4, we report NTB energy contributions, obtained at no additional cost, in comparison with the established spin-contamination-free [69,70] ALMO-EDA [69-77] DFT energy components. Orthogonalization (Ortho) and hybridization/correlation (Hyb + C) energy terms are dominant and correlate well with the respective Pauli and spin-coupling/charge-transfer terms (SC + CT) from ALMO-EDA. Differences between methods are attributed to the use of orthogonalized fragments and the inclusion of kinetic energy in ALMO-EDA. Consistent with both the earliest [78,79] and recent [68] reports, chemical bond formation is mainly driven by hybridization. The form of hybridization energy in NTB, however, is quite peculiar $(-(ab|ba))$ and can be interpreted as either the asymptotic Heitler-London resonance for quasi-orthogonal atomions (Appendix B of Ref. [43]) or a sum of inter-atomic asymptotic exchange and static correlation.

In summary, we have derived simple analytical expressions for dynamic correlation energy, the corresponding potential, and the total energy functional for the $H_2$ molecule in the framework of the independent atom ansatz of DFT. The mathematical expressions are very accurate and provide new insights into chemical bonding, such as the absence of the kinetic energy contribution and the central role of asymptotics and quasi-orthogonal atomic states.

This work was supported by National Science Foundation award number CHE-2154781.

[1] P.-O. Löwdin, Physical Review **97**, 1474 (1955).
[2] J. Hubbard, Proceedings of the Royal Society of London. Series A. Mathematical and Physical Sciences **240**, 539 (1957).
[3] O. Sinanoğlu, The Journal of Chemical Physics **36**, 706 (1962).
[4] J. Čížek, The Journal of Chemical Physics **45**, 4256 (1966).
[5] J. Cizek and J. Paldus, Physica Scripta **21**, 251 (1980).
[6] P. R. Taylor, G. Bacskay, N. Hush, and A. Hurley, Chemical Physics Letters **41**, 444 (1976).
[7] J. Pople, R. Krishnan, H. Schlegel, and J. Binkley, International Journal of Quantum Chemistry **14**, 545 (1978).
[8] R. J. Bartlett and G. D. Purvis, International Journal of Quantum Chemistry **14**, 561 (1978).
[9] G. D. Purvis and R. J. Bartlett, The Journal of Chemical Physics **76**, 1910 (1982).
[10] K. Raghavachari, G. W. Trucks, J. A. Pople, and M. Head-Gordon, Chemical Physics Letters **157**, 479 (1989).
[11] F. Neese, A. Hansen, and D. G. Liakos, The Journal of chemical physics **131** (2009).
[12] C. Riplinger and F. Neese, The Journal of chemical physics **138** (2013).
[13] S. Hirata, P.-D. Fan, A. A. Auer, M. Nooijen, and P. Piecuch, The Journal of chemical physics **121**, 12197 (2004).
[14] C. Møller and M. S. Plesset, Physical review **46**, 618 (1934).
[15] R. Krishnan, M. Frisch, and J. Pople, The Journal of Chemical Physics **72**, 4244 (1980).
[16] J. A. Pople, R. Krishnan, H. Schlegel, and J. S. Binkley, International Journal of Quantum Chemistry **16**, 225 (1979).
[17] P. Y. Ayala and G. E. Scuseria, The Journal of chemical physics **110**, 3660 (1999).
[18] S. Grimme, The Journal of chemical physics **118**, 9095 (2003).
[19] R. T. McGibbon, A. G. Taube, A. G. Donchev, K. Siva, F. Hernández, C. Hargus, K.-H. Law, J. L. Klepeis, and D. E. Shaw, The Journal of chemical physics **147** (2017).
[20] E. Wigner, Physical Review **46**, 1002 (1934).
[21] D. Ceperley, Physical Review B **18**, 3126 (1978).
[22] D. M. Ceperley and B. J. Alder, Physical review letters **45**, 566 (1980).
[23] S. H. Vosko, L. Wilk, and M. Nusair, Canadian Journal of physics **58**, 1200 (1980).
[24] P. Hohenberg and W. Kohn, Physical Review **136**, B864 (1964).
[25] W. Kohn and L. J. Sham, Physical Review **140**, A1133 (1965).
[26] M. Rasolt and D. Geldart, Physical Review Letters **35**, 1234 (1975).
[27] J. Perdew, D. Langreth, and V. Sahni, Physical Review Letters **38**, 1030 (1977).
[28] D. C. Langreth and M. Mehl, Physical Review B **28**, 1809 (1983).
[29] J. P. Perdew, Physical review B **33**, 8822 (1986).
[30] J. P. Perdew, J. A. Chevary, S. H. Vosko, K. A. Jackson, M. R. Pederson, D. J. Singh, and C. Fiolhais, Physical review B **46**, 6671 (1992).
[31] C. Lee, W. Yang, and R. G. Parr, Physical review B **37**, 785 (1988).
[32] J. P. Perdew, K. Burke, and M. Ernzerhof, Physical Review Letters **77**, 3865 (1996).
[33] J. P. Perdew, A. Ruzsinszky, G. I. Csonka, O. A. Vydrov, G. E. Scuseria, L. A. Constantin, X. Zhou, and K. Burke, Physical review letters **100**, 136406 (2008).
[34] M. Filatov and W. Thiel, Physical Review A **57**, 189 (1998).
[35] J. P. Perdew, S. Kurth, A. Zupan, and P. Blaha, Physical review letters **82**, 2544 (1999).
[36] J. Tao, J. P. Perdew, V. N. Staroverov, and G. E. Scuseria, Physical review letters **91**, 146401 (2003).
[37] J. P. Perdew, A. Ruzsinszky, G. I. Csonka, L. A. Constantin, and J. Sun, Physical Review Letters **103**, 026403 (2009).
[38] J. Sun, A. Ruzsinszky, and J. P. Perdew, Physical Review Letters **115**, 036402 (2015).

[39] Y. Zhao and D. G. Truhlar, Theoretical chemistry accounts **120**, 215 (2008).
[40] T. Schwabe and S. Grimme, Physical Chemistry Chemical Physics **9**, 3397 (2007).
[41] E. Trushin, A. Thierbach, and A. Görling, The Journal of Chemical Physics **154** (2021).
[42] J. P. Perdew and K. Schmidt, in *AIP Conference Proceedings* (American Institute of Physics, 2001), pp. 1.
[43] A. V. Mironenko, Preprint arXiv: 2204.04554v3. Under review in Physical Review B. (2024).
[44] M. Fuchs, Y. M. Niquet, X. Gonze, and K. Burke, The Journal of Chemical Physics **122**, 094116 (2005).
[45] A. J. Cohen, P. Mori-Sanchez, and W. Yang, The Journal of Chemical Physics **129**, 121104 (2008).
[46] https://cccbdb.nist.gov/.
[47] J. A. Pople, M. Head-Gordon, D. J. Fox, K. Raghavachari, and L. A. Curtiss, The Journal of Chemical Physics **90**, 5622 (1989).
[48] M. G. Medvedev, I. S. Bushmarinov, J. Sun, J. P. Perdew, and K. A. Lyssenko, Science **355**, 49 (2017).
[49] A. K. Theophilou, The Journal of Chemical Physics **149**, 074104 (2018).
[50] T. Kato, Communications on Pure and Applied Mathematics **10**, 151 (1957).
[51] A. V. Mironenko, The Journal of Physical Chemistry A **127**, 7836 (2023).
[52] A. Szabo and N. S. Ostlund, *Modern Quantum Chemistry: Introduction to Advanced Electronic Structure Theory* (Dover Publications, Mineola, New York, 1989).
[53] J. C. Slater, The Journal of Chemical Physics **1**, 687 (1933).
[54] R. F. Bader, Accounts of chemical research **18**, 9 (1985).
[55] R. P. Feynman, Physical review **56**, 340 (1939).
[56] C. Coulson, J. Lewis, S. Atom, and U. A. Viewpoints, Quantum theory **2**, 185 (1961).
[57] H. Hellmann, Zeitschrift für Physik **85**, 180 (1933).
[58] F. Hirshfeld and S. Rzotkiewicz, Molecular Physics **27**, 1319 (1974).
[59] M. Spackman and E. Maslen, The Journal of Physical Chemistry **90**, 2020 (1986).
[60] K. Ruedenberg, Reviews of Modern Physics **34**, 326 (1962).
[61] M. A. C. Nascimento, Journal of the Brazilian Chemical Society **19**, 245 (2008).
[62] M. Feinberg and K. Ruedenberg, The Journal of Chemical Physics **54**, 1495 (1971).
[63] W. A. Goddard III and C. W. Wilson Jr, Theoretica chimica acta **26**, 211 (1972).
[64] G. B. Bacskay and S. Nordholm, The Journal of Physical Chemistry A **117**, 7946 (2013).
[65] M. W. Schmidt, J. Ivanic, and K. Ruedenberg, The Journal of chemical physics **140** (2014).
[66] T. M. Cardozo and M. A. C. Nascimento, The Journal of chemical physics **130** (2009).
[67] A. C. West, M. W. Schmidt, M. S. Gordon, and K. Ruedenberg, The Journal of Physical Chemistry A **121**, 1086 (2017).
[68] D. S. Levine and M. Head-Gordon, Nature communications **11**, 4893 (2020).
[69] D. S. Levine and M. Head-Gordon, Proceedings of the National Academy of Sciences of the United States of America **114**, 12649 (2017).
[70] D. S. Levine, P. R. Horn, Y. Mao, and M. Head-Gordon, J Chem Theory Comput **12**, 4812 (2016).
[71] R. Z. Khaliullin, M. Head-Gordon, and A. T. Bell, J Chem Phys **124**, 204105 (2006).
[72] R. Z. Khaliullin, E. A. Cobar, R. C. Lochan, A. T. Bell, and M. Head-Gordon, Journal of Physical Chemistry A **111**, 8753 (2007).
[73] R. Z. Khaliullin, A. T. Bell, and M. Head-Gordon, J Chem Phys **128**, 184112 (2008).
[74] P. R. Horn, E. J. Sundstrom, T. A. Baker, and M. Head-Gordon, J Chem Phys **138**, 134119 (2013).
[75] P. R. Horn, Y. Mao, and M. Head-Gordon, J Chem Phys **144**, 114107 (2016).
[76] P. R. Horn and M. Head-Gordon, J Chem Phys **144**, 084118 (2016).
[77] P. R. Horn, Y. Mao, and M. Head-Gordon, Phys Chem Chem Phys **18**, 23067 (2016).
[78] W. Heitler and F. London, Zeitschrift für Physik **44**, 455 (1927).
[79] L. Pauling, Journal of the American Chemical Society **53**, 3225 (1931).

## Supplementary Information for

## Analytical Correlation in the H₂ Molecule from the Independent Atom Ansatz

Alanna "Lanie" Leung, Alexander V. Mironenko*

*Department of Chemical and Biomolecular Engineering,*
*University of Illinois Urbana-Champaign, Urbana, Illinois 61820*

*Email: alexmir@illinois.edu

**Section S1. Model implementation and computational details.**

The NTB binding energy of an H₂ molecule was computed as a function of the inter-nuclear distance $R$ using in-house codes that employ Libcint [1] and XCFun [2] libraries and PySCF ver. 1.7.6 software [3] to handle orbital integration. The STO-6G basis set with the exponent $\zeta$ optimized on the fly, referred to as $\zeta$-STO-6G, was used in one implementation of NTB codes. The Gaussian basis exponents $\alpha_k$ and Gaussian contraction coefficients $g_k$ were rescaled using the standard method [4,5] as follows:

$$\alpha_k = \left(\frac{\zeta}{\zeta^{ref}}\right)^2 \alpha_k^{ref}, \qquad (S1)$$

and

$$g_k = \left(\frac{\zeta}{\zeta^{ref}}\right)^{3/2} g_k^{ref}, \qquad (S2)$$

where $\zeta^{ref} = 1.24$ is the standard hydrogen exponent of the STO-NG basis set, and $\alpha_k^{ref}$ and $g_k^{ref}$ are standard STO-6G Gaussian exponents and contraction coefficients, respectively. They were taken from Ref. [6] and are tabulated in Table S1.

Table S1. Standard STO-6G hydrogen basis exponents and contraction coefficients.

| Basis exponent | Contraction coefficient |
|---|---|
| 0.3552322122E+02 | 0.9163596281E-02 |
| 0.6513143725E+01 | 0.4936149294E-01 |
| 0.1822142904E+01 | 0.1685383049E+00 |
| 0.6259552659E+00 | 0.3705627997E+00 |
| 0.2430767471E+00 | 0.4164915298E+00 |
| 0.1001124280E+00 | 0.1303340841E+00 |

A second implementation of NTB employed Slater-type orbitals that were similarly rescaled, referred to as $\zeta$-STO. Some of the expressions used in the $\zeta$-STO implementation can be found in Ref. [7] and in the source code in Section S10. The supplementary text primarily uses results from the $\zeta$-STO-6G basis, while $\zeta$-STO appears in the main text. Differences in their equilibrium binding energies, bond lengths, and frequencies are insignificant at 0.000 Å, 0.001 eV, and 3 cm⁻¹, respectively. The two bases have an average difference in binding energies of 0.59 meV for $R > 0.3$ Å and are considered to have equivalent performance.

Two methods were employed to determine $\zeta(R)$: virial and variational. Both methods rely on the integral form of the atomion equation:

$$\varepsilon_a = \left\langle \varphi_a \left| -\frac{1}{2}\nabla^2 + v_a \right| \varphi_a \right\rangle$$
$$+ \langle \varphi_a | v_b + j_b | \varphi_a \rangle$$
$$+ \left( -\frac{\sqrt{2}}{4} + \frac{1}{2} \langle \varphi_a | \varphi_b \rangle \right) (\varphi_a \varphi_b | \varphi_b \varphi_a), \quad (S3)$$

where $\varepsilon_a$ is the energy of atomion $a$ ($\varepsilon_1 = \varepsilon_2$ due to symmetry); $j_a$ is the Hartree potential due to density $\rho_a$; and $v_a$ is the corresponding nuclear potential. In the variational approach, $\varepsilon_a = \varepsilon_a(\zeta_a)$ is minimized for fixed $\zeta_b$, where $\varepsilon_a$ is the atomion energy. Then, $\varepsilon_b = \varepsilon_b(\zeta_b)$ is minimized for fixed $\zeta_a$, and steps are alternated until convergence. In the virial approach, the virial theorem is enforced by setting $\varepsilon_a = -\zeta_a^2/2$ (negative kinetic energy of an atomion) and solving the algebraic eq. (S3) for $\zeta_a$. Once $\zeta$ is found, the rest of the algorithm is nearly identical to that reported in Ref. [7]. Correlation energy $E_c^{\leftrightarrow}$ was computed using eq. (9). Promotion energy $E_{prom}$ was set to zero, based on the arguments presented in the main manuscript.

The accuracy of the NTB method was assessed by comparison of generated binding energy curves, equilibrium bond energies, equilibrium distances, and vibrational wavenumbers with DFT and full configurational interaction (FCI) data. Reference data were obtained using PySCF software and the cc-pVQZ basis set. Two nonempirical DFT functionals were considered – SCAN [8] and PBE [9]. QChem 6.0 software [10] was used to carry out ALMO-EDA energy decomposition analysis [11-19] at the $\omega$B97X-V level with HF exchange as a dispersion-free reference. Frequencies were calculated numerically near the minimum with a step size $h = 2.646 \times 10^{-3}$ Å that was optimized for PySCF DFT calculations. Experimental data were taken from Ref. [20]. Densities were computed in PySCF and visualized in VESTA software [21].

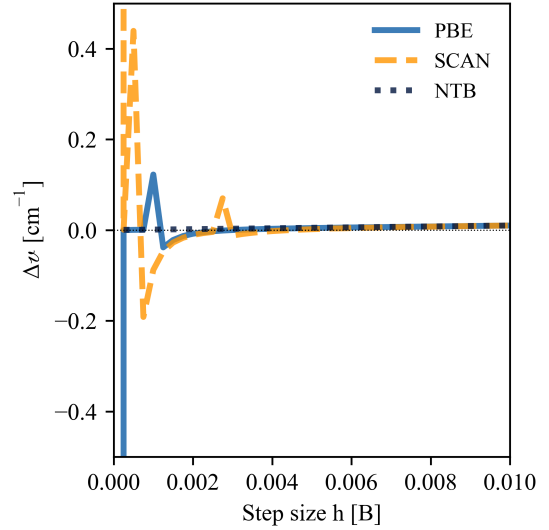

FIG S1. Changes in vibrational frequency between with respect to step size.

H$_2$ vibrational frequencies ($v$) were calculated numerically. The second derivative of total energy was approximated at the equilibrium bond distance ($R_0$) using the central difference method

$$\frac{\partial^2 E}{\partial R_0^2} = \frac{E(R_0 + h) + E(R_0 - h) - 2E(R_0)}{h^2}, \quad (S4)$$

where $h = 2.646 \times 10^{-12}$ m is step size, $R_0$ is given in meters, and $E$ is the total energy in Joules. For DFT and FCI methods, $E$ was calculated using PySCF software. For NTB, $E$ was calculated using in-house codes. $v$ was calculated as

$$v = \frac{1}{2\pi c} \sqrt{\frac{2}{m_H} \frac{\partial^2 E}{\partial r_0^2}}, \quad (S5)$$

where $c = 2.998 \times 10^{10}$ m/s is the speed of light and $m_H = 1.674 \times 10^{-27}$ kg is the mass of an H atom.

The step size $h$ was chosen after analysis of frequency as a function of step size in DFT/PBE, DFT/SCAN, and NTB. Difference in frequency between successive steps $n$ were calculated

$$\Delta v(n) = v(h_n) - v(h_{n-1}) \tag{S6}$$

for $1.0 \times 10^{-7}$ B $\leq h_n \leq 2.0 \times 10^{-2}$ B at intervals $h_n - h_{n-1} = 2.5 \times 10^{-4}$ B. For $h < 4 \times 10^{-3}$ B, the DFT frequencies are unstable (Fig. FIG S1). Both DFT/SCAN and DFT/PBE exhibit fluctuations over 0.100 cm$^{-1}$ between successive steps. For $h > 4 \times 10^{-3}$ B, all DFT and NTB $\Delta v$ increase monotonically. Frequency calculations tabulated in the manuscript Table 1 use a step size in the stable region $h = 0.005$ B.

**Section S2. Theory principles.**

In this section, we summarize the key NTB theory principles [22] that rationalize NTB equations and are applied again in this work to obtain numerically accurate mathematical forms of $E_c^{\leftrightarrow}$ and $\mu_{xc}^{\leftrightarrow,a}$.

*Principle 1.* At $R \gg 0$, there is the asymptotic correspondence between NTB and KS-DFT to $O(S)$ and between NTB and HF/CI, NTB and valence bond (VB) theories to $O(S^0)$.

*Principle 2.* At $R \gg 0$, inter-atomic integrals containing differential overlaps adopt local forms. At even larger $R$, inter-atomic exchange integrals disappear, while static correlation integrals remain.

*Principle 3.* Analytical inter-atomic terms are obtained using the method of *translatio ex infinitum.* At $R \gg 0$, NTB terms are matched with those of KS-DFT to $O(S)$, and with those of FCI to $O(S^0)$. As atoms are brought to chemically relevant $R$ values, the derived expressions are retained, provided that the local integrals are replaced with their non-local counterparts as follows:

$$-\sqrt{\pi}\langle \varphi_a | \varphi_b | \varphi_a \rangle \to -\frac{1}{4}(\varphi_a \varphi_b | \varphi_b \varphi_a). \tag{S7}$$

*Principle 4.* The orbital parity rule holds – energy terms containing an even number of atomion functions $\varphi_a$ are regarded as functionals of $\rho_a = |\varphi_a|^2$, while the terms with an odd number of $\varphi_b$ are *not* functionals of $\rho_b = |\varphi_b|^2$ and do not contribute to the atomion equation.

*Principle 5 – a void eigenpotential principle.* Energy terms obtained from eigenvalue problems (such as $E_{hyb}$) do not contribute to the atomion potential, unless off-diagonal matrix elements become zero at a finite $R$.

*Principle 6.* At $R \gg 0$, quantum fluctuations become important and are accounted by making a $\mu_{x,ab}^{R\gg 0} \to \mu_{x,ab}^{R\gg 0} + \delta$ substitution, where $\delta > 0$. There exists $R = R_\infty$, at which the $\mu_{x,ab}^{R\gg 0}$ and $\delta$ effects cancel each other identically, making $D_{ab}^x$ zero, and removing the inter-atomic exchange. In this limit, Principle 5 no longer applies, and the exchange part of $E_{hyb}$ can be differentiated and makes a contribution to $\mu_{xc}^{\leftrightarrow,a}$ (eq. (5)). The XC potential that variationally corresponds to the XC energy at $R = R_\infty$ but not at $R < R_\infty$ is referred to as *the detached potential*.

**Section S3. Comparison of local and non-local NTB.**

Binding energy and atomion equations for non-local NTB are presented in eq. (14) and (S3), respectively. For comparison, the analogous local NTB (L-NTB) expressions from Ref. [7] are

$$\Delta E(R) = 2E_{prom} + E_{es}^{\infty} + 2\sqrt{\pi}(-1 + \langle \varphi_1 | \varphi_2 \rangle)\langle \varphi_2 | \varphi_2 | \varphi_1 \rangle + \Delta E_{xc}, \tag{S8}$$

where $E_{prom}$ is taken as constant, and $\Delta E_{xc} = E_{xc}[\rho_1 + \rho_2] - 2E_{xc}[\rho_1]$ is computed using the PBE XC functional.

The integral atomion equation in L-NTB is

$$\varepsilon_a = \langle \varphi_a | -\tfrac{1}{2}\nabla^2 + v_a | \varphi_a \rangle$$
$$+ \langle \varphi_a | v_b + j_b | \varphi_a \rangle - \sqrt{\pi} \langle \varphi_a | \varphi_b | \varphi_a \rangle$$
$$+ \sqrt{\pi} \langle \varphi_a | \varphi_b \rangle \langle \varphi_b | \varphi_b | \varphi_a \rangle. \tag{S9}$$

Below, we compare local (L-NTB; Ref. [7]) and non-local (this work) flavors of NTB. Energy expressions differ conceptually in the following aspects:

(1) L-NTB uses the local (1-particle) $\langle \varphi_1 | \mu_{xc,12}^{R \gg 0} | \varphi_2 \rangle = -\sqrt{\pi}(\varphi_1 \varphi_1 \varphi_2)$ resonance integral. Non-local NTB uses the non-local (2-particle) resonance integral $-\tfrac{1}{2}(\varphi_1 \varphi_2 | \varphi_1 \varphi_2)$.

(2) $E_{xc}^{\leftrightarrow}$ in L-NTB is calculated using the PBE XC functional. In non-local NTB, it is calculated analytically as $E_c^{\leftrightarrow} = \tfrac{1-\sqrt{2}}{2}(\varphi_1 \varphi_2 | \varphi_1 \varphi_2)$.

(3) In L-NTB, the MO-XC integral is $\langle \varphi_a | \mu_{xc}^{\leftrightarrow,a} | \varphi_a \rangle = -\sqrt{\pi}(\varphi_1 \varphi_1 \varphi_2)$. In non-local NTB, we derived $\langle \varphi_a | \mu_{xc}^{\leftrightarrow,a} | \varphi_a \rangle = -\tfrac{\sqrt{2}}{4}(\varphi_a \varphi_b | \varphi_b \varphi_a)$.

(4) L-NTB requires $E_{prom} = 0.32$ eV in H$_2$ whereas non-local NTB uses $E_{prom} = 0$.

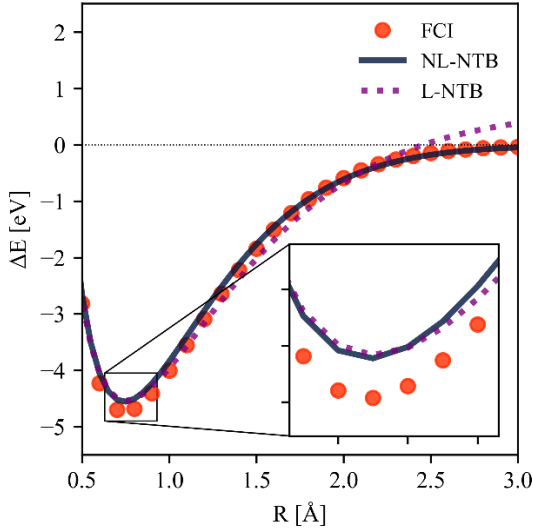

FIG S2. H$_2$ binding energies computed with L-NTB (purple dotted line), non-local NTB (blue solid line), and FCI (red circles). $E_{prom} = 0.32$ eV is used in L-NTB, similar to Ref. [7].

In Table S2, we report the performance of L- and non-local NTB at their respective equilibrium geometries using the virial and variational approaches to optimize $\zeta$. $\Delta E_0$, $R_0$, and wavenumber of non-local NTB with the virial approach are all closer to experimental values than those of their L-NTB counterparts. Virial L-NTB $E_0$, $R_0$, and $\nu$ differ from experimental values by 0.20 eV, 0.010 Å, and 359 cm$^{-1}$, compared to 0.19 eV, 0.002 Å, and 13 cm$^{-1}$ for the virial non-local NTB.

The binding energy curves of the L- and non-local NTB are qualitatively similar at $R < 1.0$ Å but diverge more at larger separations (Fig. S2). This is due to the use of the positive constant $E_{prom}$ in L-NTB, which improves energy description at the minimum (see Ref. [7]) at the expense of the correct bond dissociation, since $\Delta E(R \to \infty) = 2E_{prom}$. In contrast, non-local NTB dissociates the bond correctly due to $E_{prom} = 0$.

The superior performance of non-local NTB over the local version at chemical bonding $R$ values can be attributed to the correct asymptotic dependence of non-local resonance integrals on differential overlap. In Ref. [22], it was noted that the extension of the DFT variational principle to NTB requires the NTB total energy to be of $O(\varphi_a \varphi_b)^2$. This is the dependence exhibited by non-local integrals $D_{ab} = -\tfrac{1}{2}(\varphi_a \varphi_b | \varphi_a \varphi_b)$, while the use of local integrals $D_{ab} = -\sqrt{\pi}(\varphi_a \varphi_a \varphi_b)$ result in total energies being of $O(\varphi_a \varphi_b)$, causing overbinding, too flat energy curves, and significantly underestimated wavenumbers (Table S2). The correct quadratic dependence of non-local $D_{ab}$ on the differential overlap leads to the improved description of the H-H chemical bond across all three metrics considered.

**Table S2.** Comparison between H$_2$ non-local NTB and L-NTB using fixed $\zeta$ values, virial, and variational approaches to compute equilibrium geometries, energies, and vibrational frequencies.

| Method | Basis | XC Functional | Optimization | $R_0$ (Å) | $\Delta E_0$ (eV) | $v$ (cm$^{-1}$) |
|---|---|---|---|---|---|---|
| Local NTB [7] | $\zeta$-STO-6G | PBE | Virial | 0.751 | -4.541 | 4042 |
| Local NTB | $\zeta$-STO-6G | PBE | Variational | 0.717 | -4.068 | 3557 |
| Non-local NTB | $\zeta$-STO | Analytic | Virial | 0.743 | -4.558 | 4388 |
| Non-local NTB | $\zeta$-STO-6G | Analytic | Virial | 0.743 | -4.557 | 4385 |
| Non-local NTB | $\zeta$-STO-6G | SCAN | Virial | 0.740 | -4.654 | 4445 |
| Non-local NTB | $\zeta$-STO-6G | Analytic | Variational | 0.703 | -4.553 | 3818 |
| Non-local NTB | $\zeta$-STO-6G | Analytic | $\zeta = 1.000$ | 0.826 | -4.192 | 4129 |
| Non-local NTB | $\zeta$-STO-6G | Analytic | $\zeta = 1.089$ | 0.759 | -4.561 | 4686 |
| Experiment [20] | | | -- | 0.741 | -4.743 | 4401 |

**Section S4. Comparison between virial and variational approaches in non-local NTB.**

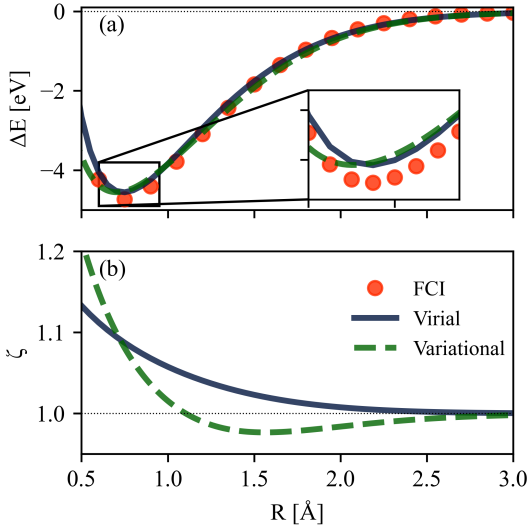

FIG S3. $\Delta E$ per atom and $\zeta$ of non-local NTB optimized by the virial approach (solid dark blue line) and variationally (green dashed line). FCI was used as a reference (red circles).

In this section, we compare $\zeta$ values, binding energies, vibrational wavenumbers, and equilibrium geometries that result from the virial and variational approaches to the atomion eigenvalue problem when using non-local resonance integrals. At large $R$, both variationally- and virially-optimized $\zeta$ approach the value for free H, $\zeta = 1.0$. Based on optimized exponents in molecules [5], it is expected that $\zeta \geq 1.0$ at chemically relevant distances in H$_2$. The virially-optimized $\zeta$ values behave conventionally by approaching $\zeta = 1.0$ from above at larger $R$. However, the variationally-optimized $\zeta$ fall below 1.0 at $R > 1.10$ Å, making atomions more diffuse than free atom orbitals in a vacuum (Fig. S3). Such a dependence bears similarity to $\zeta$ variations observed in the Heitler-London treatment of the H$_2$ molecule with $\zeta < 1$ for $R > 1.43$ Å [23].

The $R_0$, $\Delta E_0$, and $v$ predicted by the non-local NTB exhibit higher accuracy with the virial than with variational approach. In particular, the variational method yields signed errors of -0.038 Å, 0.19 eV, and -583 cm$^{-1}$. We attribute large errors in the bond length and the vibrational wavenumber to the too rapid change of $\zeta$ to too high values at smaller $R$ (Fig. S3b). Since the bond length can be regarded as an average of atomic radii [24], and the $\zeta$ value is formally equal to the inverse of an effective atomic radius, too high $\zeta$ near $R = 0.741$ Å (Fig. S3b) and below would correspond to too small $R_0$ values. Similarly, too rapidly increasing $\zeta$ at smaller $R$ would reduce the rate of increase and decrease of dominating orthogonalization and hybridization terms, respectively, rendering the potential energy too soft and the vibrational wavenumber – underestimated. Indeed, in a series "constant $\zeta = 1.089$"→"virially varying $\zeta$" →"(more rapidly) variationally varying $\zeta$", vibrational wavenumbers decrease as 4686 cm$^{-1}$→4385 cm$^{-1}$→3818 cm$^{-1}$, according to Table S2.

The reason for too rapid variations and too high values of variationally optimized $\zeta$ likely lies in the minimal basis set being too limited. The $\zeta > 1$ values arise due to the neighboring atom's potential in the atomion equation being negative, largely due to not fully screened nuclear potential $v_b$, when atomic densities overlap. Restricting the total density to the sum of spherically symmetric densities eliminates the physical density buildup between atoms (Figure 3c of the main manuscript), reducing the screening of $v_b$ and resulting in the total potential being too low and, consequently, the $\zeta$ value being too high and varying too rapidly. Inclusion of polarization functions should make the screening more efficient and reconcile virial and variational methods, making the latter more accurate.

It is interesting that in Fig. S3a, at $R < R_0$ the variational method predicts energy values considerably lower than the exact energies from the FCI, which would not be possible in the conventional electronic structure methods based on the independent electron approximation. We hypothesize that such underestimation is related to the v-representability problem in DFT [25]. The DFT variational principle, which lies at the core of the NTB method, only applies to v-representable densities, i.e., densities that arise from a ground-state wave function that can be derived from a Hamiltonian containing some potential $v$. It is evident that density of the form $\rho = |1s_a|^2 + |1s_b|^2$ is only v-representative at $R \to \infty$. At chemically bonding $R$ and large orbital overlaps, electron density at $\mathbf{r} \to \mathbf{R}_a$, where $\mathbf{R}_a$ is the position of nucleus $a$, will be anisotropic. This, however, is not physically possible, since at such $\mathbf{r}$ the potential will be dominated by isotropic $\frac{1}{|\mathbf{r}-\mathbf{R}_a|} \to -\infty$, and all other potential terms will be negligibly small in magnitude relative to it. Consequently, electron density in the vicinity of $\mathbf{R}_a$ can only be isotropic. Additionally, in the interatomic region, it is difficult to rationalize $v = v_1 + v_2$ that would yield the above $\rho$, as the presence of $v_2$ in the vicinity of $\rho_1$ should perturb it and break the spherical symmetry, not just rescale $\zeta$. Since non-$v$-representative densities are employed in the variational method of NTB, energy can take values lower than the true variational energy. This problem can be solved by employing larger basis sets, but not all basis sets may lead to $v$-representable densities. The problem of $v$-representability does not arise in the Kohn-Sham DFT, since all allowable minimal-basis densities are essentially identical for Hartree-Fock and DFT.

It is notable that the limited minimal basis set does not represent a problem for the virial method, where the virial theorem may act as a beneficial physical constraint that ensures the correct asymptotic density decay $\rho_a \sim \exp(-2(-2\varepsilon_a)^{1/2} r)$ [25], seemingly important for the theory derived in the asymptotic large separation limit. Since the virial theorem approach results in more accurate $\Delta E_0$, $R_0$, and frequency for non-local NTB, all the results presented in this work are obtained using the virial approach to optimize $\zeta$.

**Section S5. Optimization of $\zeta$.**

The sensitivity of $\Delta E$ to $\zeta$ is further investigated by comparison of non-local NTB $\Delta E$ curves generated using virially-optimized and static $\zeta$ (Fig. S4). In the static model, $\zeta$ are fixed to values of 1.000 and 1.089, which are their free atom and equilibrium values, respectively. The $R_0$, $\Delta E_0$, and frequency are reported in Table 2. In the $\zeta = 1.000$ case, the overly diffuse atomions predict excessively weak bonds at small $R$ and increased equilibrium bond length. The $\Delta E(\zeta = 1.089)$ curve under-binds slightly at both small and moderately large interatomic distances relative to $R_0$. Differences in errors in $R_0$, $\Delta E_0$, and frequency for the fixed $\zeta = 1.089$ case are 0.012 Å larger, 0.004 eV smaller, and 269 cm$^{-1}$ larger than those in virially-optimized NTB, respectively. For calculation of equilibrium geometry and energies, a fixed $\zeta$ model is a reasonable alternative to solving the atomion equation at each inter-atomic distance. To achieve the highest frequency accuracy in the NTB framework, $\zeta$ must be adapted to the atomic environment.

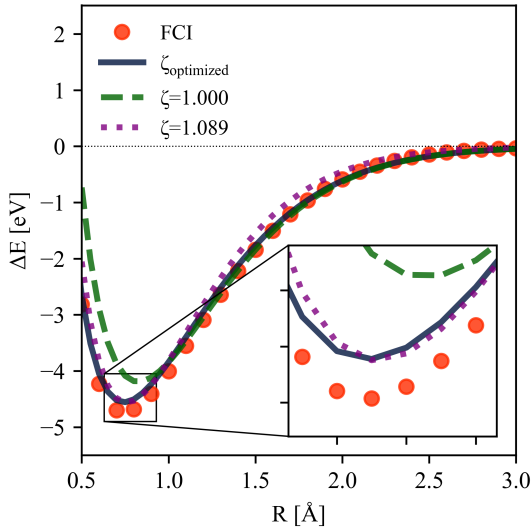

FIG S4. Non-local NTB binding energies with $\zeta$ optimized at every geometry using the virial theorem (dark blue solid line) and $\zeta$ fixed to its free atom value, $\zeta = 1.000$ (green dashed line) and NTB equilibrium values, $\zeta = 1.089$ (purple dotted line). FCI was used as a reference (red circles).

### Section S6. Sensitivity of energy to the $\mu_{xc}^{\leftrightarrow}$ prefactor

In this section, we characterize the performance of variants of the inter-atomic XC potential integral obtained by altering the derivation in the main text. The derivation of the integral $\langle \varphi_1 | \mu_{xc}^{\leftrightarrow,1} | \varphi_1 \rangle = -\frac{\sqrt{2}}{4}(\varphi_1 \varphi_2 | \varphi_2 \varphi_1)$ in eq. (12) requires applications of Principles summarized in Section S2 before differentiating

$$\mu_{xc}^{\leftrightarrow,a} = \frac{\delta(D_{ab}^{sc} + D_{ba}^{sc} + \sqrt{2}(D_{ab}^{x} + D_{ba}^{x}))}{\delta \rho_a}. \tag{S10}$$

Enforcing the void eigenvalue principle and orbital parity rule prevent $D_{ab}^{sc} + D_{ba}^{sc}$ and $D_{ba}^{x}$, respectively, from contributing to the potential. Neglect of either or both conditions during the derivation of the potential results in different prefactors $k$ for the integral $\langle \varphi_1 | \mu_{xc}^{\leftrightarrow,1} | \varphi_1 \rangle = k(\varphi_1 \varphi_2 | \varphi_2 \varphi_1)$, which are summarized in Table S3.

Table S3. Variants of $\mu_{xc}^{\leftrightarrow}$ prefactors, principles neglected to derive them, terms differentiated, equilibrium bond lengths and binding energies predicted, along with orbital exponents at the minimum.

| $\langle \varphi_1 \| \mu_{xc}^{\leftrightarrow,1} \| \varphi_1 \rangle$ prefactor expression | Numeric prefactor value | Principles neglected | Terms differentiated | $R_0$ (Å) | $\Delta E_0$ (eV) | $\zeta_0$ |
|---|---|---|---|---|---|---|
| $\frac{(1-\sqrt{2})}{4}$ | -0.104 | Void eigenvalue | $D_{ab}^{sc}, D_{ab}^{x}$ | 0.810 | -4.257 | 1.015 |
| $\frac{(1-\sqrt{2})}{2}$ | -0.207 | Void eigenvalue, orbital parity | $D_{ab}^{sc}, D_{ba}^{sc}, D_{ab}^{x}, D_{ba}^{x}$ | 0.781 | -4.380 | 1.045 |
| $-\frac{\sqrt{2}}{4}$ | -0.354 | None | $D_{ab}^{x}$ | 0.743 | -4.557 | 1.089 |
| $-\frac{\sqrt{2}}{2}$ | -0.707 | Orbital parity | $D_{ab}^{x}, D_{ba}^{x}$ | 0.663 | -4.991 | 1.196 |

The NTB $\Delta E$ curves are shown to be sensitive to the choice of the inter-atomic XC potential (Fig. S5). More negative prefactors decrease $R_0$, make $\Delta E_0$ more negative, and increase the equilibrium $\zeta$ ($\zeta_0$). To satisfy the virial theorem, more negative $\langle \varphi_1 | \mu_{xc}^{\leftrightarrow,1} | \varphi_1 \rangle$ contributions are compensated by atomion contraction. The lower differential overlap at small $R$ reduces the magnitude of repulsive contributions, particularly $E_{ortho}$. Thus, more negative prefactors tend to lower both $R_0$ and $\Delta E_0$. Of the variants for inter-atomic XC potential integrals alternatively derived, $\langle \varphi_1 | \mu_{xc}^{\leftrightarrow,1} | \varphi_1 \rangle = -\frac{\sqrt{2}}{4} (\varphi_1 \varphi_2 | \varphi_2 \varphi_1)$ predicts the most accurate equilibrium energy and bond lengths.

**Section S7. NTB performance for $H_2(T)$.**

In this section, we report the performance of non-local NTB for triplet $H_2(T)$. As both bonding and antibonding orbitals are singly occupied in $H_2(T)$, $E_{hyb} = 0$. Since no electron excitations are possible within the states spanned by the atomion basis, $E_c^{\leftrightarrow} = 0$ and $\mu_{xc}^{\leftrightarrow} = 0$. Therefore, the $H_2(T)$ binding energy is a sum of the remaining energy contributions,

$$\Delta E = E_{ortho} + \Delta E_{es}. \tag{S11}$$

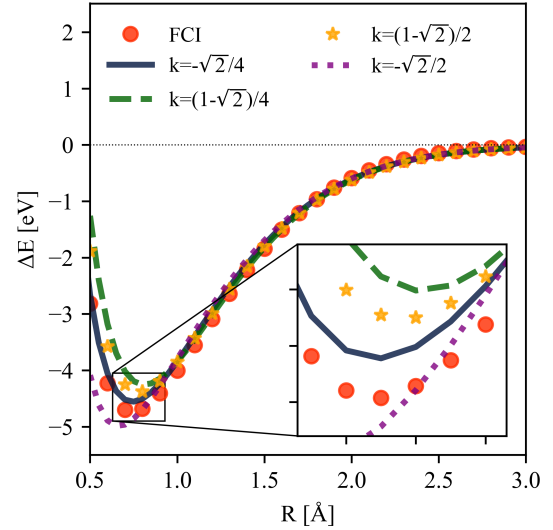

FIG S5. Non-local NTB energies using $\langle \varphi_1 | \mu_{xc}^{\leftrightarrow,1} | \varphi_1 \rangle = k(\varphi_1 \varphi_2 | \varphi_2 \varphi_1)$. The model standard $k = -\frac{\sqrt{2}}{4}$ (dark blue solid line), $\frac{(1-\sqrt{2})}{4}$ (green dashed line), $\frac{(1-\sqrt{2})}{2}$ (yellow stars), and $-\frac{\sqrt{2}}{2}$ (purple dotted line) are plotted with FCI as a reference.

An open question is whether there is an additional exchange interaction, associated with same-spin electrons. We confirm that inter-atomic exchange contributions are confined to $D_{ab}$ by comparing $\Delta E$ curves to which an explicit $E_x^{\leftrightarrow}$, computed using the SCAN functional, has been artificially added. In Figure S6, the $H_2(T)$ energy curves are reported. The explicit exchange term leads to overbinding relative to the reference. Exclusion of $E_x^{\leftrightarrow}$ confirms that all inter-atomic exchange is accounted for by the energy contributions $E_{ortho}$ and $E_{hyb}$ in NTB, and no extra terms are present.

As the $H_2(T)$ NTB curve deviates from FCI, the question remains about physical effects not captured by the model. As we have demonstrated in the main manuscript, in $H_2(S)$ the sum of symmetric densities deviates considerably from the FCI density in the inter-nuclear region. To obtain more accurate densities, polarization functions are required, which, however, should not considerably affect energy, in light of the first Theophilou theorem [26]. Since this theorem applies to the ground state, while $H_2(T)$ is the excited state, we can expect that polarization functions would have a more significant effect on H$_2$(T) energy. In fact, this is confirmed in Figure S7. In particular, in the atomion $|\varphi_a\rangle$ that solves the atomion equation $\varepsilon_a|\varphi_a\rangle = \left[-\frac{1}{2}\nabla^2 + v_a + v_b + j_b - P_b\mu_{x,ba}^{R\gg0}\right]|\varphi_a\rangle$, which, unlike the one for H$_2$(S), does not contain the correlation potential, the polarization function will work to deplete electron density in the interatomic region, in order to reduce $j_b$ and $-P_b\mu_{x,ba}^{R\gg0}|\varphi_a\rangle$ and thereby minimize $\varepsilon_a$. Consequently, densities would overlap less, and $\Delta E_{es}$ would become less negative. To model this effect semi-quantitatively, we have set $\Delta E_{es} = 0$ in eq. (S11), resulting in the closest-lying non-local NTB energy curve relative to the reference (Fig. S6).

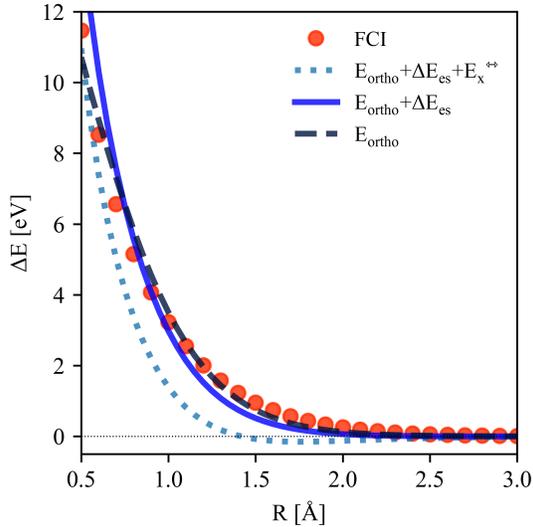

FIG S6. Non-local NTB energies for $H_2(T)$. $E_{ortho} + \Delta E_{es} + E_x^{\leftrightarrow}$ (light blue dotted line), $E_{ortho} + \Delta E_{es}$ (blue solid line), and $E_{ortho}$ (dark blue dashed line).

**Section S8. Asymptotic NTB/VB correspondence and reconciliation of local and non-local theories.**

In the main text, the asymptotic correspondence between NTB and VB theories has been posited as a rationale for the CI treatment of dynamic correlation. However, it can also be noted that the derived expression $E_c^{\leftrightarrow} = \Delta_{NTB} - (\Delta_{NTB}^2 + K_{NTB}^2)^{\frac{1}{2}}$ is not correspondent with the asymptotic VB. Specifically, the equality $\Delta_{NTB} = K_{NTB} = -0.5(\varphi_1\varphi_2|\varphi_2\varphi_1)$ prevents $E_c^{\leftrightarrow}$ from approaching $-0.5(\varphi_1\varphi_2|\varphi_2\varphi_1)$ in order to yield its VB counterpart $-(\varphi_1\varphi_2|\varphi_2\varphi_1)$, when summed with the static correlation part of $E_{hyb}$. Instead, the NTB/VB correspondence requires that $\Delta_{NTB}$ goes to zero more rapidly than $K_{NTB}$. In this context, we demonstrate that the faster $\Delta_{NTB}$ decay can be achieved by incorporating the zero-point energy of a vacuum arising from quantum fluctuations.

According to the Heisenberg uncertainty principle $\Delta\mathbf{r}\Delta\mathbf{p}\sim\hbar^3$, the momentum and thus the kinetic energy of a finite empty space (such as the Universe) is non-zero due to quantum fluctuations. For the external potential $v_{ext} = v_1 + v_2$, we argue that vacuum fluctuations can be incorporated by redefining zero potential through making the following substitution:

$$v_{ext} \rightarrow v_{ext} + \delta_v, \qquad (S12)$$

where $\delta_v > 0$. The constant shift in the external potential affects $\Delta_{NTB}$ and $E_{hyb}$, but not $K_{NTB}$. To observe its effect, we first note that $\Delta_{NTB} = 0.5(\varepsilon_2^{mo} - \varepsilon_1^{mo})$. At large $R$, NTB and KS-DFT are correspondent, and $\varepsilon_i^{mo}$ (or perhaps its exchange-only part) shall coincide with the corresponding KS energies $\varepsilon_i$ (Section VE of Ref. [22]). Since $\varepsilon_i = \langle\psi_i|H_{ks}|\psi_i\rangle$, for $H_{ks} \rightarrow H_{ks} + \delta_v$, the bonding $\varepsilon_i$ and thus $\varepsilon_i^{mo}$ will be shifted upward, whereas the antibonding $\varepsilon_i^{mo}$ – downward. More specifically, by repeating the derivation of the atomion equation from KS equations containing the $\delta_v$ shift (Section VD of Ref. [22]), we find that $\delta_v$ affects $D_{ab}$ integrals in much the same way as electron quantum fluctuations $\delta$:

$$\mu_{x,ab}^{R\gg0} \rightarrow \mu_{x,ab}^{R\gg0} + \delta + \delta_v. \qquad (S13)$$

Evidently, the zero-point energy of the vacuum does not affect electron correlation. From this it follows that $K_{NTB}$ is independent of $\delta_v$, which also ensures that the $\Delta_{NTB} \rightarrow \Delta_{NTB} - 2\delta_v$ shift cancels out in the $E_c^{\leftrightarrow}$ expression.

After taking into account the vacuum energy shift, we have the following set of equations:

$$\begin{aligned} E_{x-hyb} + E_c^{\leftrightarrow} &= -(\Delta_{NTB}^2 + K_{NTB}^2)^{\frac{1}{2}} \\ \Delta_{NTB} &= -2\langle\varphi_a|\mu_{x,ab}^{R\gg0} + \delta + \delta_v|\varphi_b\rangle \\ K_{NTB} &= 2\langle\varphi_a|\mu_{x,ab}^{R\gg0} + \delta|\varphi_b\rangle \end{aligned} \qquad (S14)$$

From eq. (S14) it follows that $\Delta_{NTB}$ is always slightly less than $K_{NTB}$ in magnitude by the amount $2\delta_v$ times overlap. Therefore, at large $R$, when the effect of $\mu_{x,ab}^{R\gg0}$ is canceled by $\delta + \delta_v$ and $\Delta_{NTB} = 0$, $K_{NTB} \neq 0$ and $E_{x-hyb} + E_c^{\leftrightarrow} = E_c^{\leftrightarrow} = -K_{NTB} = -0.5(\varphi_1\varphi_2|\varphi_2\varphi_1) + O(\delta)$. When added to $E_{c-hyb}$, the asymptotic

inter-atomic correlation energy of the VB theory is correctly reproduced, ensuring asymptotic equivalence. The $R$ value at which $\mu_{x,ab}^{R\gg 0} = -(\delta + \delta_v)$ is denoted by $R_L$, and we refer to this value as the *near asymptotic limit*, in contrast to the far asymptotic limit at $R = R_\infty$ ($R_\infty > R_L$), where $\mu_{x,ab}^{R\gg 0} = -\delta$.

It is instructive to determine the form of $\mu_{xc}^{\leftrightarrow,a}$ in the near asymptotic limit at $R = R_L$. First of all, we note that since $K_{NTB} \neq 0$, the term derived from $E_c^{\leftrightarrow}$ does not appear in $\mu_{xc}^{\leftrightarrow,a}$. Since $D_{ab}^x = 0$, the exchange term does appear in $\mu_{xc}^{\leftrightarrow,a}$. By using the argument from Section VIIC of Ref. [22] and after replacing non-local integrals with their local forms, we find that $\mu_{xc}^{\leftrightarrow,a} = \mu_x^{\leftrightarrow,a} = -\sqrt{\pi}\varphi_b$, which is the XC potential form used in the L-NTB theory variant in Ref. [7]. It can also be shown that the local integral form $D_{ab} = -\sqrt{\pi}(aab)$ arises in the same limit, since $E_{hyb} = E_{c-hyb}$. We arrive at the conclusion that the L-NTB formalism arises when self-consistency is enforced at $R = R_L < R_\infty$ and local integral forms are applied at chemical bonding distances (along with the local XC functional).

It should be noted that L-NTB equations have been derived in Ref. [22] in an alternative way, using the same static correlation (SC-) interpretation of $E_{hyb}$ but without accounting for quantum fluctuations. It is important that the derivation in the $R_L$ limit, presented here, is free of conceptual limitations of the original SC-interpretation (see Section IXA of Ref. [27]) while yielding essentially the same equations. In particular, it correctly identifies $E_{xc}^{\leftrightarrow} = E_c^{\leftrightarrow}$ as the only contribution, whereas in the original form, we had $E_{xc}^{\leftrightarrow} = E_x^{\leftrightarrow}$, much like in a standard semilocal KS DFT theory with dominant exchange effects.

In the $R_\infty$-limit, $D_{ab}^x$, $\Delta_{NTB}$, and $K_{NTB}$ are all zero, making $E_c^{\leftrightarrow}$ differentiable and leading to the form of $\mu_{xc}^{\leftrightarrow,a}$ shown in eq. (11). The non-local NTB theory, introduced in this work, is obtained when the self-consistency is enforced at $R = R_\infty$, and local integrals are replaced with their non-local counterparts at chemically bonding $R$.

We conclude this section by summarizing the hierarchy of interatomic distances and interatomic overlaps ($RS$-hierarchy) used in the NTB theory in Table S4.

Table S4.

| $RS$-hierarchy | Property | NTB theory features derived in the limit |
|---|---|---|
| $O(S)$ | NTB/KS-DFT correspondence | Atomion equation, $E$ functional |
| $O(S^0)$ | NTB/KS-DFT/HF-CI/VB correspondence | Non-local $D_{ab}$ |
| Quasi-classical[27] | Local inter-atomic integrals | Local $D_{ab}$, 1-particle CI for $E_c^{\leftrightarrow}$, pairwise $\mu_{xc,ab}^{R\gg 0}$ |
| Near-asymptotic ($R_L$) | Electron localization, exchange disappearance; static correlation dominates | $K_{NTB}$ in $E_c^{\leftrightarrow}$, $D_{ab}$ and $\mu_{xc}^{\leftrightarrow,a}$ in L-NTB |
| Far-asymptotic ($R_\infty$) | $E_c^{\leftrightarrow}$ disappearance, Debye-Hückel limit[27] | $\mu_{xc}^{\leftrightarrow,a}$ in NTB after L→NL transition |

**Section S9. Estimate of NTB errors due to polarization and the simplicity of the analytical functional form.**

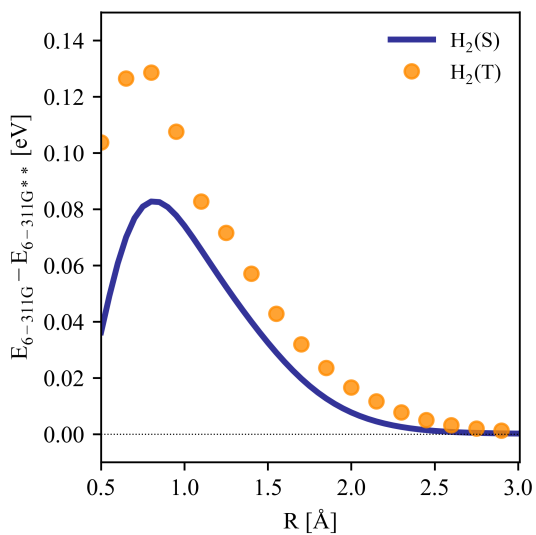

FIG S7. Differences in $H_2(S)$ and $H_2(T)$ energies calculated by 6-311G and 6-311G** basis sets.

In the main text, the lack of polarization functions in atomions and the simplicity of the analytical expression were highlighted as factors contributing to the underestimation of the H-H bond energy by 0.19 eV with the NTB method. To estimate the effect of polarization on energy in a semi-quantitative manner, we compare energy differences between 6-311G (approximately equivalent to $\zeta$-STO) and 6-311G** basis sets at the Kohn-Sham DFT/SCAN level of theory (Fig. S7). At R= 0.741 Å, the difference is found to be 0.081 eV. The error due to neglect of 1s-2s, 1s-2p, etc. interactions in the analytical dynamic correlation expression can be estimated by the sum of two differences: between analytical and SCAN XC energies for $\rho = \rho_1 + \rho_2$ (Fig. 2) and between SCAN and FCI XC energies for the FCI density (cc-pVQZ basis set). At $R = 0.741$ Å, SCAN and FCI XC energies differ by 0.0088 eV, indicating that the SCAN XC energy is nearly exact. SCAN and analytical XC energies, in turn, differ by 0.098 eV. The total contribution of all factors is approximately 0.19 eV, which is equal to the difference between NTB and experimental values (0.19 eV; Table 1).

**Section S10. Implementation of NTB.**
The $\zeta$-STO implementation of non-local NTB with the virial approach in Python 3 can be followed below.

```python
import numpy as np
from scipy.linalg import eigh
from scipy.optimize import fsolve, minimize
from scipy.special import exp1 as E1

S_mat = np.zeros((2,2))
D_mat = np.zeros((2,2))

def get_abba(R,Z):
    '''Compute (ab|ba) integral'''
    w = Z*R
    gamma = np.euler_gamma
    S = np.exp(-w) * (1 + w + w**2/3)
    S_prime = np.exp(w) * (1 - w + w**2/3)
    abba = Z/5 * (-np.exp(-2*w)*(-25/8 + 23*w/4 + 3*w**2 + w**3/3) \
        + 6/w *(S**2 * (gamma + np.log(w)) - S_prime**2*E1(4*w) +
2*S*S_prime*E1(2*w)))
    return abba

def get_S_matrix(R,Z):
    '''Calculate overlap matrix off-diagonal elements'''
    S_ab = ((R**2*Z**2)/3 + R*Z + 1)*np.exp(-R*Z)
    S_mat[0,1] = S_ab
```

```python
    S_mat[1,0] = S_ab

def get_D_matrix(R,Z):
    '''Calculate Huckel matrix'''
    D_mat[0,0] = -(Z**2)/2
    D_mat[1,1] = -(Z**2)/2
    Dij = -1/2 * get_abba(R,Z)
    D_mat[0,1] = Dij
    D_mat[1,0] = Dij

def get_e_ortho():
    '''Calculate resonance energy'''
    e_orthogonal = - (S_mat[0,1] * D_mat[1,0] + S_mat[1,0] * D_mat[0,1])
    return e_orthogonal

def get_e_hybridization():
    D_eigenvalues, D_eigenvectors = eigh(D_mat)
    eps = D_eigenvalues[0]
    e_hyb = (2 * eps) - (D_mat[0,0] + D_mat[1,1])
    return e_hyb

def get_e_electrostatics(R,Z):
    '''Calculate energy from external potentials and Coulomb'''
    e_NN = 1/R
    e_Ne = Z * np.exp(-2*Z*R) * (1 + (1/(Z*R))) - (1/R)
    e_ee = (1/R) - (Z * np.exp(-2*Z*R) * ((1/(Z*R)) + (11/8) + (3*Z*R/4) + (((Z*R)**2)/6)))
    e_es = e_NN + (2 * e_Ne) + e_ee
    return e_es

def get_e_c(R,Z):
    '''Calculate correlation energy'''
    abba = get_abba(R,Z)
    e_c = (1-np.sqrt(2))/2 * abba
    return e_c

def atomion_eqn(Z,R):
    '''Generalized Anderson equation'''
    get_S_matrix(R,Z)
    get_D_matrix(R,Z)
    e_Ne = Z * np.exp(-2*Z*R) * (1 + (1/(Z*R))) - (1/R)
    e_ee = 1/R - Z * np.exp(-2*Z*R) * ((1/(Z*R)) + (11/8) + (3*Z*R/4) + (((Z*R)**2)/6))
    Daba = -1/2 * get_abba(R,Z)/np.sqrt(2)
    residual = Z**2 - Z + (e_Ne + e_ee) + Daba - S_mat[0,1] * D_mat[1,0]
```

```python
        return residual

def get_singleval(R, output='single'):
    '''
    Calculates H2 bond formation energy in Hartrees for a single interatomic
distance

    Args:
        R (float): bond distance in Bohr
        output (str): whether to return energy contributions and zeta or only
binding energy

    Returns:
        Z_val (float): optimal zeta for given geometry
        e_ortho (float): energy required to orthogonalize basis in Hartree
        e_hyb (float): resonance energy in Hartree
        e_es (float): electrostatic energies in Hartree
        e_c (float): inter-atomic dynamic correlation energy in Hartree
        e_tot (float): binding energy in Hartree
    '''
    Z_val = fsolve(atomion_eqn, x0=1.0, args=(R))

    e_ortho = get_e_ortho()
    e_hyb = get_e_hybridization()
    e_es = get_e_electrostatics(R, Z_val)
    e_c = get_e_c(R, Z_val)

    e_tot = e_ortho + e_hyb + e_es + e_c

    if output == 'single':
        output = e_tot
    else:
        output = [Z_val, e_ortho, e_hyb, e_es, e_c, e_tot]
    return output
```

**Section S11. Discussion on the lack of the kinetic energy contribution in the NTB binding energy expression.**

The binding energy (BE) expression in the NTB model (eq. (14)) does not include the kinetic energy (KE) term, which seems to contradict the virial theorem. This contradiction becomes even more remarkable when considering the successful application of the virial theorem in solving the atomion equation within this work. In other words, while KE plays a crucial role in determining the optimal value of $\zeta_0$, it does not appear in calculations of binding energy $\Delta E$.

As a test, we compared the "would-be" KE associated with BE ($+13.6 \times 2 + 4.56$=31.76 eV or 1.1672 Hartree) and the KE present in the atomion equation ($2 \times \zeta_0^2/2$ for $\zeta_0 = 1.089$, which is 1.1859 Hartree). Both values are reasonably close, differing by about 0.50 eV. While it appears that the majority of the virial kinetic energy is associated with the KE of atomions and atomic contraction, the atomion KE does not contribute to $\Delta E$ due to the atomion renormalization phenomenon ($E_{prom} = 0$).

In Chapter VC of the second version of the arXiv preprint [27], one of us introduced the collapsing pilot wave interpretation of quantum mechanics, which may provide an explanation for the absence of KE. In this framework, a wave and a particle are distinct entities with separate energetic characteristics, where the wave guides the motion of the particle. This allows for a particle to have an invariant KE independently of the KE of the wavefunction, which would cancel out in BE calculations.

The development of the collapsing pilot wave interpretation was driven by the desire to reconcile and explain seemingly unorthodox (and numerically validated) features of the NTB theory. The main corollary of this interpretation is the asymptotic quasi-classicality, which implies that at large distances from a nucleus, an electron moves quasi-classically, even for principal quantum numbers as low as 1. Quasi-classicality implies locality [28], which, in turn, can explain the following NTB model features:

1) The importance and ubiquity of asymptotically local inter-atomic integrals in NTB;
2) The fact that the asymptotic $\mu_{xc}^{\leftrightarrow,a}$ resembles the Debye-Hückel correlation potential for the classical electron gas [29];
3) The appearance of $O(S)$ terms in the BE expression that are consistent with the one-particle (mean-field) molecular orbital theory, instead of $O(S^2)$ terms that would follow from the asymptotic correspondence with the valence bond theory;
4) The one-particle CI description of the dynamic correlation energy in H$_2$ (eq. (9));
5) Disappearance of exchange in the quasi-classical limit, since exchange is a quantum, not classical, phenomenon;
6) Electron density distribution being additive ($\rho = \rho_1 + \rho_2$) and not affected by wave interference;
7) The importance and ubiquity of a "constant orbital substitution trick" in deriving local resonance integrals from non-local ones: $\langle \varphi_e | \mu_{xc,ba}^{R \gg 0} | \varphi_a \rangle \to \langle \varphi_e | \mu_{xc,ba}^{R \gg 0} \rangle \varphi_a^\infty$, which follows from the asymptotic discretization of the wavefunction [27], a corollary to the collapsing pilot wave interpretation;
8) The fact that integrals of the form $\langle \varphi_b | \mu_{x,ba}^{R \gg 0} + \delta | \varphi_a \rangle$ approach zero at $R \to \infty$ and do not become positive, which is the consequence of the asymptotic wavefunction discretization (see #7 above);
9) Asymptotic pairwise $\mu_{xc,ba}^{R \gg 0}$ for degenerate atomic states [27].

The electron KE invariance and cancellation in the BE expression are further supported by two pieces of evidence. First is the fact that $E_{prom} = 0$, such that an electron in $\varphi_a$ and $\varphi_a^0$ has the same kinetic (and potential) energy. Second comes from the comparison of the asymptotically local inter-atomic potential $\mu_{xc,ba}^{R \gg 0} = \sqrt{\pi} \varphi_b = \sqrt{\pi \rho_b}$ and the Debye-Hückel correlation potential $\sqrt{\frac{\pi}{T}} \rho$, where $T$ is the temperature of a classical electron plasma. Proceeding purely formally, we recognize that in statistical mechanics $T$ is proportional to $KE$, and thus any variations in $KE$ would lead to variations in the prefactor in $\mu_{xc,ba}^{R \gg 0}$. However, one of us has shown [7] that the prefactor remains constant and equal to $\sqrt{\pi}$ for various $R$ and $x$ values in $H_x$ clusters. Thus, KE must remain constant as well.

Having summarized the evidence for the KE invariance, the question remains about the mechanism for keeping KE constant. To this end, the following inexact, qualitative picture can be proposed, motivated by the idea that the presence of T in the expressions above implies a system interacting with a macroscopic (infinite-dimensional) environment. The wavefunction collapse is associated with a projection from the infinitely-dimensional Hilbert space onto one of its dimensions, such as from $|\varphi\rangle = \sum_i c_i |g_i\rangle$ onto $|g_i\rangle$. Since $\varphi(\mathbf{r}) = \int \varphi(\mathbf{r}') \delta(\mathbf{r} - \mathbf{r}') d\mathbf{r}'$, where $\varphi$ is an atomion, we can formally regard $\delta(\mathbf{r} - \mathbf{r}')$ and thus every $\mathbf{r}'$ as associated with different dimensions. We alternatively write the "master" atomion wavefunction in infinite dimensions as $\Psi(\{\mathbf{r}_i'\}|\{\varphi_i\}) = \varphi_1 \psi(\mathbf{r}_1') + \varphi_2 \psi(\mathbf{r}_2') + \cdots$, where $\varphi_i = \varphi(\mathbf{r}_i)$ with $\mathbf{r}_i$ being points in the 3D space, and $\mathbf{r}_1'$, $\mathbf{r}_2'$, etc. formally corresponding to orthogonal coordinate axes in the infinitely-dimensional Cartesian space (hidden dimensions), so that $\psi(\mathbf{r}_1')$ and $\psi(\mathbf{r}_2')$ are orthogonal. It is easy to confirm that $\Psi$ is normalized. The probability of finding an electron at $\mathbf{r}_1$ can then be determined as $\langle\Psi|\psi(\mathbf{r}_1')\rangle\langle\psi(\mathbf{r}_1')|\Psi\rangle = |\varphi(\mathbf{r}_1)|^2$. As the wavefunction responds to the environment, coefficients $\varphi_1, \varphi_2$, etc. change, whereas $\psi$ is assumed constant (akin to the amplitude modulation). Then, if energy is computed for the Hamiltonian expressed in terms of $\mathbf{r}$ (i.e. in terms of the guiding wave $\varphi(\mathbf{r})$), $\langle\Psi|H|\Psi\rangle = \langle\varphi|H|\varphi\rangle = E[\varphi]$. However, if energy is computed for the Hamiltonian in terms of $\{\mathbf{r}_i'\}$ (i.e., for an infinite-dimensional "master" wavefunction that collapses to a particle upon nucleus-electron interaction), $\langle\Psi|H|\Psi\rangle = \langle\psi|H|\psi\rangle = const$. The overall scenario resembles interaction of a system and a heat bath in thermostatted molecular dynamics simulations.

The above qualitative picture would be consistent with NTB if $\langle\psi|H|\psi\rangle$ can related to the energy of a free H atom ($-1/2$), and if the correct effective temperature $T$ can be extracted from it. To this end, Svidzinsky, Herschbach et al. [30-34] have shown that the dimensional rescaling of the Schrödinger equation for H to an infinite number of dimensions reproduces the Bohr model of an H atom. Notably, in the Bohr model an electron moves classically (consistent with quasiclassicality in NTB) and has one degree of freedom. From statistical mechanics we know that $KE = \frac{1}{2}T$ for such a case, and thus $T = 1$ for $KE = 1/2$. This is the $T$ value required to reproduce the numerically correct prefactor in $\mu_{xc,ba}^{R \gg 0}$ from the Debye-Hückel potential (*vide supra*).

The analysis above is evidently very superficial and far from complete. A more detailed quantitative investigation would be necessary to further prove or disprove the arguments made.

As a final note, we reconcile the negligible role of KE in binding described by NTB and the fact that it plays a crucial role in KS-DFT. To this end, we note that both NTB and KS-DFT are formally exact, and have the following total energy functionals:

$$E_{KS}[\rho] = T_{KS}[\rho] + E_{es}[\rho] + E_{xc}[\rho],$$
$$E_{NTB}[\rho] = T_{NTB}[\rho] + E_{hyb} + E_{ortho} + E_{es}[\rho] + E_{xc}[\rho]. \quad (S15)$$

A comparison of equations reveals the formal correspondence $T_{KS}[\rho] = T_{NTB}[\rho] + E_{hyb} + E_{ortho}$. Even when $T_{NTB}$ is constant and cancels out in BE calculations, $T_{KS}$ will not be zero and will formally contain $E_{hyb}$ and $E_{ortho}$. Since $E_{hyb}$ is negative for chemical bonds and $E_{ortho}$ is positive, the $T_{KS}$ curve would resemble a typical binding energy curve with a minimum, repulsive, and attractive branches, the behavior observed previously [35]. Including integrals of the kinetic energy operator into $E_{ortho}$ and $E_{hyb}$ (as in ALMO-EDA) will make $T_{KS}$ too negative, necessitating the addition of the atomic contraction term into $T_{NTB}$.

**References.**

[1]     Q. Sun, Journal of Computational Chemistry **36**, 1664 (2015).
[2]     U. Ekstrom, L. Visscher, R. Bast, A. J. Thorvaldsen, and K. Ruud, J Chem Theory Comput **6**, 1971 (2010).
[3]     Q. Sun *et al.*, J Chem Phys **153**, 024109 (2020).
[4]     A. Szabo and N. S. Ostlund, *Modern Quantum Chemistry: Introduction to Advanced Electronic Structure Theory* (Dover Publications, Mineola, New York, 1989).
[5]     W. J. Hehre, R. F. Stewart, and J. A. Pople, The Journal of Chemical Physics **51**, 2657 (1969).
[6]     B. P. Pritchard, D. Altarawy, B. Didier, T. D. Gibson, and T. L. Windus, J Chem Inf Model **59**, 4814 (2019).
[7]     A. V. Mironenko, J Phys Chem A **127**, 7836 (2023).
[8]     J. Sun, A. Ruzsinszky, and J. P. Perdew, Phys Rev Lett **115**, 036402 (2015).
[9]     J. P. Perdew, K. Burke, and M. Ernzerhof, Phys Rev Lett **77**, 3865 (1996).
[10]    Y. Shao *et al.*, Molecular Physics **113**, 184 (2014).

[11]     R. Z. Khaliullin, M. Head-Gordon, and A. T. Bell, J Chem Phys **124**, 204105 (2006).
[12]     R. Z. Khaliullin, E. A. Cobar, R. C. Lochan, A. T. Bell, and M. Head-Gordon, Journal of Physical Chemistry A **111**, 8753 (2007).
[13]     R. Z. Khaliullin, A. T. Bell, and M. Head-Gordon, J Chem Phys **128**, 184112 (2008).
[14]     P. R. Horn, E. J. Sundstrom, T. A. Baker, and M. Head-Gordon, J Chem Phys **138**, 134119 (2013).
[15]     P. R. Horn, Y. Mao, and M. Head-Gordon, J Chem Phys **144**, 114107 (2016).
[16]     P. R. Horn and M. Head-Gordon, J Chem Phys **144**, 084118 (2016).
[17]     D. S. Levine, P. R. Horn, Y. Mao, and M. Head-Gordon, J Chem Theory Comput **12**, 4812 (2016).
[18]     P. R. Horn, Y. Mao, and M. Head-Gordon, Phys Chem Chem Phys **18**, 23067 (2016).
[19]     D. S. Levine and M. Head-Gordon, Proceedings of the National Academy of Sciences of the United States of America **114**, 12649 (2017).
[20]     https://cccbdb.nist.gov/.
[21]     J. D. Chai and M. Head-Gordon, J Chem Phys **128**, 084106 (2008).
[22]     A. V. Mironenko, Preprint arXiv: 2204.04554v3  (2024).
[23]     N. Rosen, Physical Review **38**, 2099 (1931).
[24]     L. Pauling, Proceedings of the National Academy of Sciences **18**, 293 (1932).
[25]     R. G. Parr and W. Yang, Annual review of physical chemistry **46**, 701 (1995).
[26]     A. K. Theophilou, Journal of Chemical Physics **149** (2018).
[27]     A. V. Mironenko, Preprint arXiv: 2204.04554v2  (2024).
[28]     L. D. Landau and E. M. Lifshitz, *Quantum mechanics: non-relativistic theory* (Elsevier, 2013), Vol. 3.
[29]     L. D. Landau and E. M. Lifshitz, *Statistical Physics: Volume 5* (Elsevier, 2013), Vol. 5.
[30]     D. Frantz and D. Herschbach, Chemical physics **126**, 59 (1988).
[31]     A. Svidzinsky, G. Chen, S. Chin, M. Kim, D. Ma, R. Murawski, A. Sergeev, M. Scully, and D. Herschbach, International Reviews in Physical Chemistry **27**, 665 (2008).
[32]     A. A. Svidzinsky, M. O. Scully, and D. R. Herschbach, Physical review letters **95**, 080401 (2005).
[33]     A. A. Svidzinsky, M. O. Scully, and D. R. Herschbach, Proceedings of the National Academy of Sciences **102**, 11985 (2005).
[34]     A. Trabesinger, Nature Physics  (2005).
[35]     D. S. Levine and M. Head-Gordon, Nature communications **11**, 4893 (2020).